# Can Excitonic Mechanism Contribute Significantly to Superconducting Pairing in Quasi 1-D Systems?

Soumi Roy Chowdhury [a]

and

Ranjan Chaudhury [b]

S.N. Bose National Centre for Basic Sciences, Block JD, Sector III, Salt Lake, Calcutta 700 098, India

Electronic Adress:

[a] soumi@bose.res.in, soumiroychowdhury@gmail.com

[b] ranjan@bose.res.in, ranjan_021258@yahoo.com

**Abstract:**

We have revisited Cooper's one pair problem of calculating the binding energy for two electrons, experiencing an attractive interaction near the Fermi surface, in case of a quasi one dimensional lattice system. Though it is a generalized formalism, we have chosen two materials viz. $(TMTSF)_2ClO_4$ and $(TMTSF)_2AsF_6$ as examples. Some of their electronic and optical features are used for the calculation and the validity of this formalism is checked too for these materials. It is generally believed that spin density fluctuation mechanism is a very strong candidate for the occurrence of superconductivity in these quasi one dimensional materials. We have attempted to invoke another type of electronic mechanism viz. charge transfer exciton mechanism to explore if some substantial contribution for attractive coupling can be generated. This mechanism may be plausible in these types of organic materials where ionicity and co-valency co-exist. Making use of the experimental results for optical absorption spectra corresponding to the above materials, we extracted the relevant parameters for incorporating in the binding energy calculation. Our calculation shows that fermionic pair formation is possible for all the three cases viz. (1) less than half filled, (2) half filled and (3) more than half filled.

# Introduction :

In 1957 Bardeen, Cooper and Schrieffer (BCS) [1] formulated a microscopic theory of superconductivity in which attractive interaction between the electrons can bring about a new phase of a metallic system in which the ground state of whole electron system is separated from the excited state by an energy gap given by

$$\Delta(0) = 1.76 k T_C = 2k\theta_c \exp(-1/UN(E_F)).$$

Where $T_c$ is the superconducting transition temperature, $\theta_c$ is the characteristic temperature equivalent of the energy of the boson mediating the attractive pairing interaction, $N(E_F)$ is single electron density of states at the Fermi surface and U is the magnitude of the attractive interaction between the two electrons.

From the above equation it is clear that the possibility of raising the superconducting transition temperature exists if the lattice is made more polarizable i.e. if U becomes higher and $\theta_c$ is enhanced. Following these a lot of sources for exchange forces such as plasmon, spin fluctuation, exciton were proposed. We would like to examine the possibility of exciton mechanism mediating the pairing, motivated by the following facts:-

Little [2] in 1964 had suggested first the substitution of phonon mechanism by exciton mechanism in some restricted cases. The following requirements must be satisfied for the exciton mechanism to play any role:-

1) A molecular system composed of densely packed and highly polarizable molecules
2) A high concentration of conduction electrons along the system axis of transport
3) A specific character that should facilitate an effortless electron transport

The above features are found to be satisfied in many quasi one dimensional organic systems.

Almost immediately after that Toyazawa in 1966 [3] presented an one dimensional toy model containing three ions having an anion in the middle and two cations at the boundaries. An exciton is created on the central ion after an excited electron goes to conduction band, leaving a hole in the valence band. The hole and the electron are attracted through the square well potential extending over these three ions. The relative motion of the electron and hole is restricted over the nearest neighbor ions only. The picture is quite appropriate for quasi one dimensional organic superconductors of $(TMTSF)_2 X$ type having three ions, viz two cations and an anion, in an unit cell.

In recent times (1987) Varma et al [4] proposed that charge transfer exciton mechanism is a very promising candidate for bringing about superconductivity in a conducting system if there is a strong presence of ionicity besides covalency. The quasi one dimensional organic materials are generally composed of planar anion radicals (eg $ClO_4^-$, $AsF_6^-$) and cation radicals (eg. $TMTSF^+$, $BEDT-TTF^+$). Besides, the organic molecular chains are capable of orienting themselves parallel due to the π bonding between the electrons.

In this communication we would like to examine theoretically the possibility of occurrence of superconductivity in quasi one dimensional system via a non phonon mechanism. In particular we focus on some of the organic superconductors and investigate if charge transfer exciton can contribute significantly to a pairing interaction when the band structure effects are taken into account explicitly.

## Mathematical formulation:

As is well known, superconductivity doesn't occur in one dimensional material (according to Mermin Wagner theorem [5]), however Cooper pairing [6] between two electrons may still take place. Is this pairing at all possible in one dimensional lattice system? If it is so, can electronic mechanism of excitonic type be a feasible one for this system? – these are the questions we would like to address here. In particular, we are interested in calculating the binding energy (pairing energy) of the electron pair as a function of the strength of the attractive interaction, when the electron pair has zero centre of mass momentum. Our methodology is based on tight binding lattice consisting of cations and anions, with the attractive interaction extracted from the features in the optical conductivity data.

For the electrons, considered to be paired in continuum limit, the Schroedinger equation would be

$$-\hbar^2/2m \, (\nabla_1^2 + \nabla_2^2) \, \Phi(\vec{r}_1 - \vec{r}_2) + V(\vec{r}_1, \vec{r}_2)() \, \Phi(\vec{r}_1 - \vec{r}_2) = E \, \Phi(\vec{r}_1 - \vec{r}_2) \ldots\ldots\ldots\ldots(1)$$

$\Phi(\vec{r}_1 - \vec{r}_2)$ is the spin singlet pair wave function in relative co ordinate space;

$\Phi(\vec{r}_1 - \vec{r}_2) = \sum_{\vec{k}} a_{\vec{k}} \, e^{i\vec{k}.\vec{r}_1} e^{-i\vec{k}.\vec{r}_2}$. Since the exponential form can be thought of as a pair of single particle states of momentum $\vec{k}$ and $-\vec{k}$, it is clear that Cooper pair wave function is a superposition over various definite pair of single particle states ($\vec{k}\uparrow, -\vec{k}\downarrow$) with $a_{\vec{k}}$ being the corresponding probability amplitude.

In the present case of the electrons in the 1D lattice system equation (1) takes the form of:-

$$a_{\vec{k}} \left[ \left( \epsilon_0 - 2t \cos(ka) \right) + \left( \epsilon_0 - 2t \cos(-ka) \right) \right] + \sum_{\vec{k}} a_{\vec{k}} \, V_{\vec{k},\vec{k}'} = E a_{\vec{k}} \qquad (2)$$

leading to

$$2\{\epsilon_0 - 2t \cos(ka)\} \, a_{\vec{k}} - E a_{\vec{k}} = \sum_{\vec{k}} a_{\vec{k}} \, U/L \qquad (3)$$

Where E is the two particle energy eigen value and $V_{\vec{k},\vec{k'}}$ is Fourier transform of the contact potential $-U\delta(\vec{r}_1 - \vec{r}_2)$ and is equal to $-U/L$ only within the small region beyond the Fermi points where the pairing would take place (usual Cooper's model) and $\epsilon_k = (\epsilon_0 - 2t\cos(ka))$ is the single particle band energy; $L$ is the size of the 1D system in consideration and $L \to \infty$ for a macroscopic system.

Simplifying eq (3) one obtains

$$a_{\vec{k}} = \sum_{\vec{k}} \frac{\sum_{\vec{k}} a_{\vec{k}} U/L}{[2(\epsilon_0 - 2t\cos(ka)) - E]} \qquad (4)$$

Going over to the continuum limit,

$$1 = U/L \int_{E_F}^{E_F + \hbar\omega_{exciton}} \frac{N(\epsilon)d\epsilon}{2\epsilon - E} \qquad (5)$$

Here $\hbar\omega_{exciton}$ is the characteristic energy of the exciton mediating the pairing interaction and represents the range of the attraction in the energy space. The integration has been expressed in energy space with the electronic density of states (DOS) kept inside the integration owing to the fact that the range where the pairing takes place around the Fermi surface is quite large, as the exciton energy can be comparable to the Fermi energy.

As $N(\epsilon) = \frac{L}{2\pi} \frac{1}{2at\sqrt{1 - (\frac{\epsilon - \epsilon_0}{2t})^2}} \qquad (6)$

(a is lattice constant of the concerned material)

Therefore the final equation is

$$1 = U/L \frac{L}{2\pi} \int_{E_F}^{E_F + \hbar\omega_{exciton}} \frac{d\epsilon}{2at\sqrt{1 - (\frac{\epsilon - \epsilon_0}{2t})^2}} \cdot \frac{1}{(2\epsilon - E)} \qquad (7)$$

Now the dimension of $U\delta(\vec{r}_1 - \vec{r}_2)$ is energy and $\delta(\vec{r}_1 - \vec{r}_2)$ has the dimension of 1/length. Therefore the dimension of U is energy× length. Dimension of $N(\epsilon)$ is [energy]$^{-1}$. Thus the coupling constant, which is defined as the product of $V_{\vec{k},\vec{k'}}$ averaged over the Fermi surface and $N(\epsilon)$, is dimension less, as expected.

$$U = 1/[\frac{1}{2\pi} \int_{E_F}^{E_F + \hbar\omega_{exciton}} \frac{d\epsilon}{2at\sqrt{1 - (\frac{\epsilon - \epsilon_0}{2t})^2}} \cdot \frac{1}{(2\epsilon - E)}] \qquad (8)$$

The well known relationship between |W|, the pairing energy, and E is - |W|=E-2E$_F$. Here for convenience we have shifted the zero of energy to the bottom of the band that is at $(\epsilon_0 - 2t)$. In

that case for half filled band 2t is the new Fermi energy with respect to the new zero i.e, $E_F= \varepsilon_0 - \{\varepsilon - \varepsilon_0\} = 2t$

Following Cooper The upper limit of integration is kept at $E_F + \hbar\omega_{exciton}$ and the lower one at $E_F$ in equations (5)-(8).

In the next part of the calculation some optical properties of the relevant materials are studied to extract $\omega_{exciton}$. The value of plasma frequency, effective mass and scattering rate of electrons are determined via theoretical approach using some standard modelling equations. The value of excitonic energy is extracted from this part and incorporated in the binding energy calculation.

An extra hump besides Drude peak in the optical spectra graph is observed in some of the materials where ionicity and covalency are both present. $(TMTSF)_2 ClO_4$ is an ionic compound as one electron is shared between two $TMTSF^+$ ions and one $ClO_4^-$ ion [7]. After checking the nature of the graph of spectral weight for $(TMTSF)_2 ClO_4$ the standard way to estimate the total spectral weight has been made use of. The first expression is for Drude peak $\sigma_D(\omega)$ and the another is for non –Drude contribution. Besides the important constraint viz, the f-sum rule will have to be obeyed (see equation (11) below)

$$\sigma_D(\omega) = (\omega^2_{pD}/4\pi)\Gamma_D / (\omega^2 + \Gamma^2_D) \qquad (9)$$

$$\sigma_0(\omega) = (\omega^2_{po}/4\pi)\omega^2\Gamma_0 / [(\omega^2 - \omega^2_0)^2 + \omega^2\Gamma^2_0] \qquad (10)$$

$$\int\sigma_D(\omega,)d\omega + \int\sigma_0(\omega)d\omega = \omega_p^2/8 \qquad (11)$$

Furthermore the quantity $\omega_{pd}$ acts as the plasma frequency if there would be no damped harmonic oscillator like behaviour and $\Gamma_D$ acts as the corresponding scattering rate. The quantities $\omega_{po}$ and $\Gamma_0$ play the roles for the non Drude optical part like $\omega_{pd}$ and $\Gamma_D$ in the Drude part. $\omega_0$ is the excitonic frequency and $\omega_p$ is the total plasma frequency which is unique for each material. Thus $\omega_0$ is to be identified with $\omega_{exciton}$ used in the pairing calculation.

## Calculation and Results:

Making use of equations (9) and (10) we generate a curve as close as possible to the experimentally obtained graph for optical absorption [8].

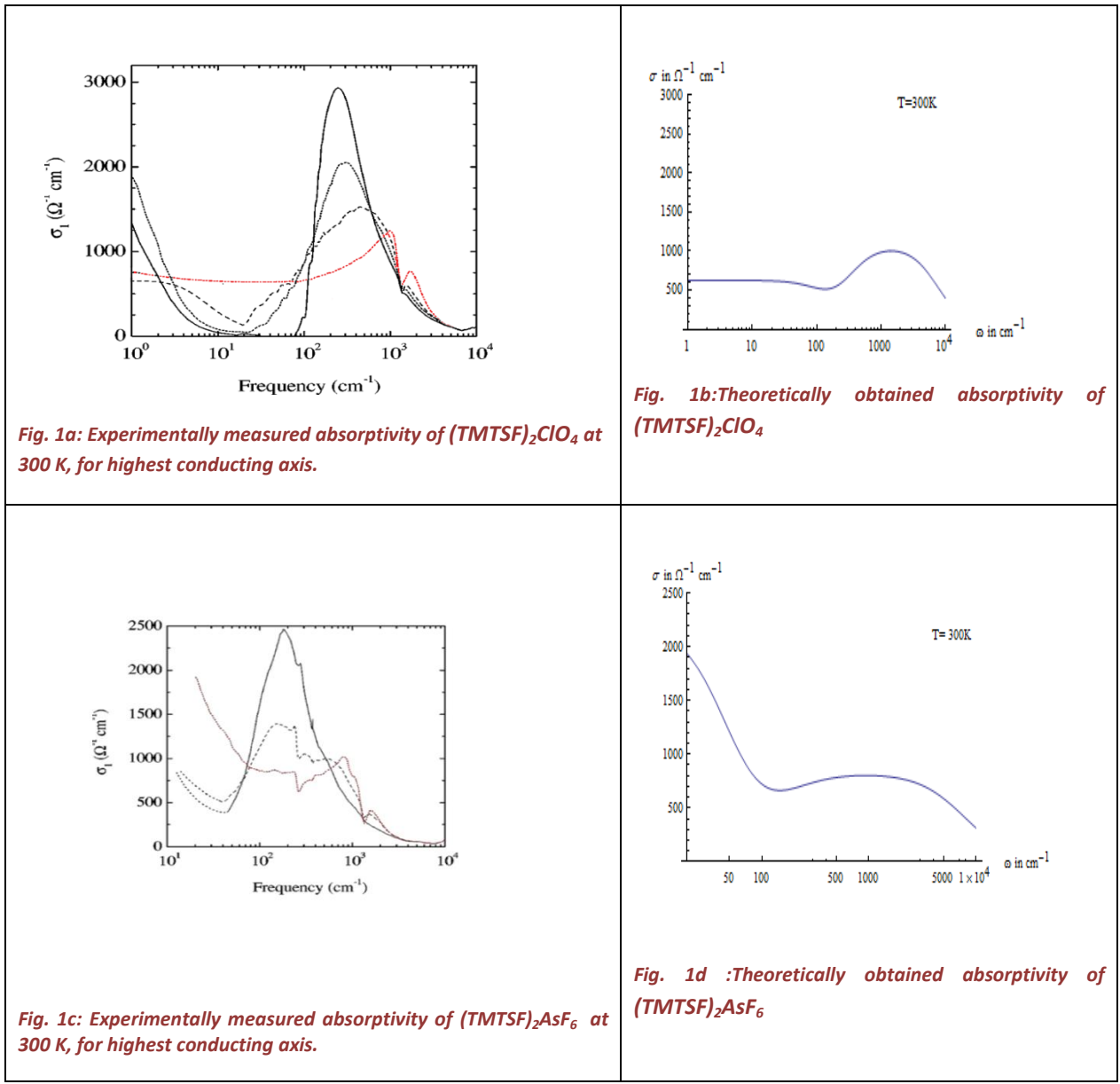

Figure 1: Comparison between experimentally observed absorptivity (red graphs in Fig. 1a and Fig. 1c)[8] and that obtained from theoretical calculation (Fig. 1b and Fig 1d).

From the above curve fitting for $(TMTSF)_2 AsF_6$ the values of $\omega_{pd}^2$ and $\omega_{po}^2$ came out to be 1400000 cm$^{-2}$ and 80000000 cm$^{-2}$ respectively; $\Gamma_0$ is 8000 cm$^{-1}$ and $\Gamma_D$ is 50 cm$^{-1}$. The excitonic energy value came out as 1050 cm$^{-1}$ whereas the original maxima of frequency in the experimentally obtained curve was 900 cm$^{-1}$. The plasma frequency is 9022 cm$^{-1}$ which is very close to experimentally obtained value 9900 cm$^{-1}$ of the material [9].

For $(TMTSF)_2ClO_4$ these values are respectively 1100000 cm$^{-2}$, 10000000 cm$^{-2}$, 8000 cm$^{-1}$ and 140 cm$^{-1}$. The excitonic energy value is 1500 cm$^{-1}$ whereas the original maxima of frequency in the experimentally obtained curve was 1000 cm$^{-1}$ The plasma frequency is 10055 cm$^{-1}$, which is close to experimentally obtained value 11000 cm$^{-1}$ [9].

| Name of the material | Experimentally obtained plasma frequency [9] | Theoretically obtained plasma frequency |
|---|---|---|
| (TMTSF)$_2$ ClO$_4$ | 11000 cm$^{-1}$ | 10055 cm$^{-1}$ |
| (TMTSF)$_2$ AsF$_6$ | 9900 cm$^{-1}$ | 9022 cm$^{-1}$ |

As stated previously this value of $\omega_0$ is used in the upper limit of the integration ($\omega_{exciton}$ of $E_F + \hbar\omega_{exciton}$) in the attractive interaction energy and pairing energy calculation. Different values of U are taken to generate different values of pairing energy and then it is multiplied by $N(\epsilon_F)$ for that concerned filling to determine the coupling constant. Then the graph between coupling constant and pairing energy is drawn.

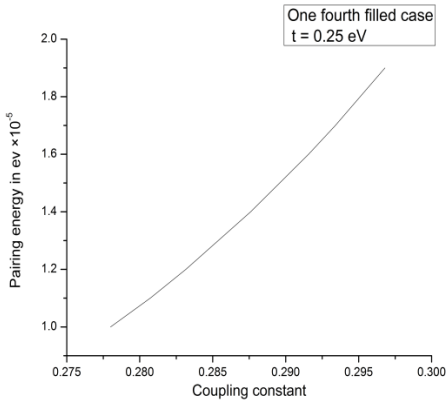

Fig. 2a : ¼ filled case

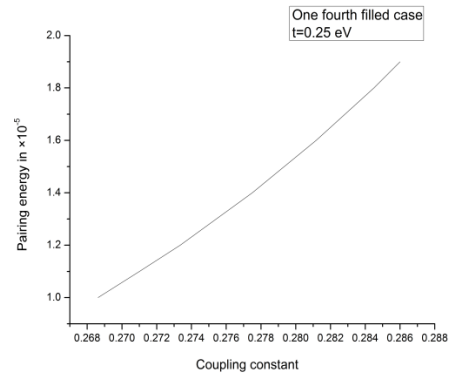

Fig. 3a: ¼ filled case

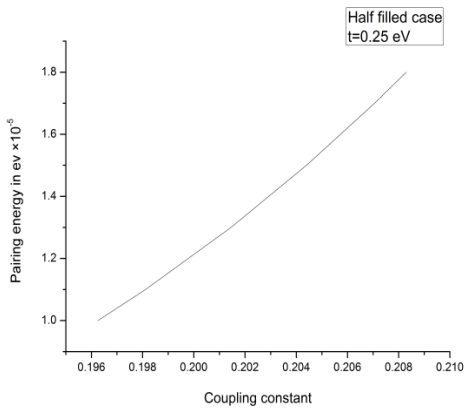

Fig. 2b: ½ filled case

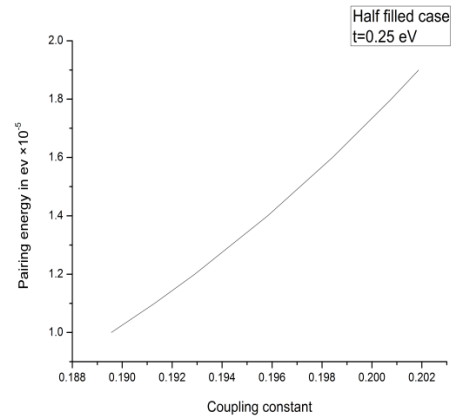

Fig. 3b: ½ filled case

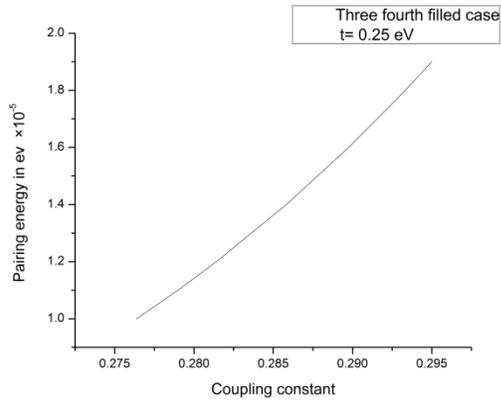 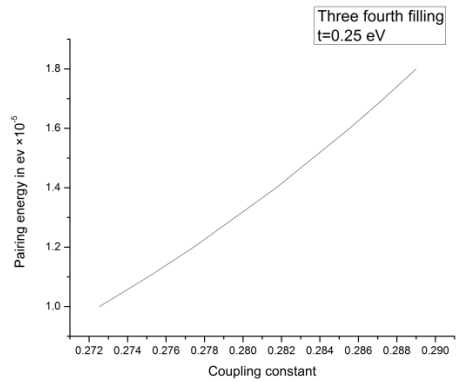

*Fig. 2c: ¾ filled case*  *Fig. 3c: ¾ filled case*

*Figure 2: Plot of pairing energy vs. coupling constant for ( TMTSF)$_2$AsF$_6$*

*Figure 3: Plot of pairing energy vs. coupling constant for (TMTSF)$_2$ClO$_4$*

For illustration we now take a particular example. The half filled case of (TMTSF)$_2$ ClO$_4$ material the value of U/L , corresponding to the binding energy of .00001 ev is taken. Recalling equation (6) we have

$$N(\epsilon_F) = \frac{L}{2\pi} \frac{1}{2at} \text{ with } \epsilon_F = \epsilon_0 .$$

Then the coupling constant becomes 0.19

Thus the coupling constant in this case belongs to the weak coupling regime and is independent of the lattice constant.

## Conclusion and Discussion:

Our main conclusions are the following:-

(i) Pairing for any filling in 1D lattice is possible.

(ii) Pairing can take place with arbitrarily small attractive interaction, as in Cooper case for continuum.

(iii) The pairing mechanism based on charge transfer exciton exchange looks quite plausible for quasi 1-D organic materials.

(iv) The pairing in more than half filled case has been studied using particles (electrons); however physically it is more appropriate to use the picture of hole-hole pairing in this case.

More elaborate calculation for superconductivity taking into account the inter-chain processes will be taken up along the line of BCS [1] approach soon as has been done for some of the high $T_C$ layered materials earlier[10].

## Acknowledgement:

One of the authors (SRC), would like to acknowledge the help of Subhajit Sarkar and Sumit ghosh regarding computational work.

## References:

1. J.Bardeen, L.N.Cooper, J.R.Schrieffer, Phys. Rev. **108** 1175 1957

2. W.A.Little, Phys.Rev.**134**  A1416 (1964)

3. Y.Toyazawa, M.Inoue, T.Inui et al, J.Phys.Soc.Japan **21**  208  (1966)

4. C.M.Varma, Schmitt-Rink, Elihu Abrahams, Solid State Communication **62** 681 (1987)

5. D.Mermin, H.Wagner,  Phys. Rev. Lett. **17**  1133  (1966).

6. Leon N. Cooper  Phys. Rev. **104**  1189  (1956)

7. T.Ishiguro, K. Yamaji, G.Saito, Organic Super-Conductors (Springer, Berlin, 1998) 2$^{nd}$ Edition; D. Jérome, A. Mazaud, M. Ribault, K. Bechgaard, J. Physique Lett. **41** 95 (1980); K. Bechgaard, C.S. Jacobsen, K. Mortensen,   H.J. Pedersen, N. Thorup, Solid State Communications **33**  1119 (1980).

8. A. Schwartz, M. Dressel, and G. Gruner; Physical Review B  **58**  1261 (1998)

9.  Andrzej Graja, Low Dimensional Organic Conductors (World Scientific,1992) English Edition.

10. Ranjan chaudhury  arXiv: 0901.1438v1 (2009)